# Ultrafast Laser Inscription of a 121-Waveguide Fan-Out for Astrophotonics

R. R. Thomson[1*], R. J. Harris[2], T. A. Birks[3], G. Brown[1], J. Allington-Smith[2] and J. Bland-Hawthorn[4,5]

[1]*Scottish Universities Physics Alliance (SUPA), School of Engineering and Physical Sciences, Department of Physics, Heriot-Watt University, Riccarton Campus, Edinburgh, EH14 4AS, UK*
[2]*Centre for Advanced Instrumentation, Physics Department, Durham University, Durham DH1 3LE, UK*
[3]*Department of Physics, University of Bath, Claverton Down, Bath, BA2 7AY, UK*
[4]*Institute of Photonics and Optical Science, School of Physics, University of Sydney, NSW 2006, Australia*
[5]*Sydney Institute for Astronomy, School of Physics, University of Sydney, NSW 2006, Australia*
*Corresponding author: R.R.Thomson@hw.ac.uk*

Using ultrafast laser inscription, we report the fabrication of a prototype three-dimensional 121-waveguide fan-out device capable of reformatting the output of a 120 core multicore fiber (MCF) into a one-dimensional linear array. When used in conjunction with an actual MCF, we demonstrate that the reformatting function using this prototype would result in an overall throughput loss of $\approx$ 7.0 dB. However, if perfect coupling from the MCF into the fan-out could be achieved, the reformatting function would result in an overall loss of only $\approx$ 1.7 dB. With adequate development, similar devices could efficiently reformat the output of so-called "photonic lanterns" fabricated using highly multicore fibers.

Astrophotonics is the field where photonic principles are applied to instrumentation for astronomy. The aim of this field is to reduce the cost, size and weight of instruments, while increasing their functionality and performance. Astrophotonic technologies currently under development include integrated beam combiners for stellar interferometry [1] and photonic spectrometers for applications in multi-object spectroscopy [2].

One area where astrophotonics may make a significant impact is in the suppression of the fluorescence lines at around $\approx$1.5 μm generated by OH-radicals in the upper atmosphere [3]. Observations in this spectral region (the astronomical J- and H-bands) are of key interest to many areas of modern astronomy e.g. high-redshift observations for studies of dark matter and dark energy.

Several OH-line suppression methods have been explored, including the use of interference filters [4] and gratings masks [5], but each of these suffers from severe drawbacks. Currently, the most promising OH-line suppression route is using complex single mode (SM) fiber Bragg-gratings (FBGs) [6, 7]. Through careful design, such FBGs can, in principle, reflect >100 individual lines at high resolution ($R = \lambda/\Delta\lambda \approx 10,000$) [8], while preserving the transmission of the dark inter-line continuum of interest.

At first sight, however, there is a problem with the astrophotonic approach to OH-line suppression; in order to achieve acceptable coupling efficiencies from the telescope into an optical fiber it is necessary to use highly multimode (MM) fibers which are not suitable for FBG applications. To facilitate the use of FBGs for OH-line suppression, it is therefore necessary to couple the light from the MM fiber into SM-FBGs. However, the brightness theorem (fundamentally the second law of thermodynamics) prevents the efficient coupling of light from an arbitrarily excited MM fiber into a single SM fiber. To address this issue, the so-called photonic lantern (PL) was developed [9-12] – a remarkable guided wave transition which facilitates the efficient adiabatic coupling of light between a MM fiber (or waveguide) and an array of SM fibers (or waveguides).

Initially, PLs were fabricated using stack-and-draw techniques [9, 10] which are labor intensive and expensive – a considerable drawback if PLs are to be mass-produced for instruments on future Extremely Large Telescopes (ELTs). Recently, however, efficient PLs have also been demonstrated using two additional routes; the first using three-dimensional (3D) ultrafast laser inscription [11], the second using tapered multicore fibers (MCFs) consisting of many SM guiding cores – such as the 120-core MCF shown in Fig. 1(a) [12].

Although the ultrafast laser inscription route is particularly promising for certain applications, including high mode count PLs and PLs for regions of the spectrum where silica is unsuitable, the MCF route has already been shown to facilitate low-loss lantern transitions involving > 100 modes. Unfortunately, however, the SM's generated by the MCF-lantern are necessarily arranged into a two-dimensional array, which is undesirable for many applications. For example, the PIMMS instrument concept [13] is being developed for applications in multi-object spectroscopy. In this case, a PL is used to couple light from the telescope into a planar arrayed waveguide grating for spectral analysis. It is clear, therefore, that a photonic interconnect is required which is capable of reformatting the two-dimensional array of SM's generated by the MCF-PL into a linear array of SM's.

As a step towards properly addressing the interconnect issue, we have fabricated a prototype 3D integrated optical waveguide fan-out device - a conceptual sketch of which is shown in Fig. 2. The axes shown in Fig. 1(b) and Fig. 2 are consistent, and will be referred to in the rest of the paper. The fan-out consists of a 3D network of 121 optical waveguides. At one end, the waveguides are arranged to match the core geometry of the 120 core MCF shown in Fig. 1(a) (with an additional spare core). At the opposite end, the cores are arranged into a linear array with an inter-waveguide spacing of 50 µm.

The fan-out was fabricated using ultrafast laser inscription (ULI) [11]. Structures were inscribed using an ultrafast Yb-doped fiber laser system (IMRA FCPA µ–Jewel D400). The system supplied 340 fs pulses of 1047 nm radiation at a pulse repetition frequency of 500 kHz. The substrate material (Corning EAGLE2000) was mounted on x-y-z air–bearing translation stages (Aerotech ABL1000) which facilitated the smooth and precise translation of the sample through the laser focus. The polarization of the laser beam was adjusted to be circular and was focused below the surface of the substrate material using an aspheric lens with a numerical aperture of 0.6. All waveguides were inscribed using a translation velocity of 8.0 mm.s$^{-1}$ and 210 nJ pulses. The cross section of the waveguides was controlled using the well-known multiscan technique [11]. Consequently, each waveguide was fabricated using 20 scans of the material through the laser focus, with each scan offset from the previous scan by 0.4 µm in the x-axis.

As shown in Fig. 1(b), the MCF end of the fan-out consisted of 23 columns of waveguides. The paths of the waveguides were controlled such that adjacent columns became adjacent sections of the linear array, and such that adjacent waveguides in each column remained adjacent in the linear array. Waveguides further up each column (more positive z-axis position) were moved into lower numbered positions in the linear array (see Fig. 2). Starting from the linear array end of the fan-out, each waveguide first moved into its final z-axis position, by moving only in the z-y plane, and then into its final x-axis position, by moving only in the x-y plane. These movements were performed using two connected S-bends, each of which was formed using two arcs of 40 mm radius. For all of the waveguides, the S-bends were initiated at the same y-axis position on the sample. After inscription, the fan-out ends were ground and polished to expose the waveguide facets.

The fan-out was characterized using 1.55 µm light. In all cases the light was injected into the fan-out using a Corning SMF-28 fiber butt-coupled to the waveguide facet. Initially, the light was coupled into the MCF coupling end of the fan-out. Near field imaging indicated that the waveguides were strongly coupled, and that they each supported a single transverse mode with 1/e$^2$ mode field diameters (MFDs) of 11 ± 0.1 µm and 7 ± 0.1 µm in the x- and z-axis respectively. The strong coupling prevented us from measuring the MFDs of the waveguides at the MCF coupling end of the fan-out.

To measure the performance of the fan-out waveguides quantitatively, light was coupled into each waveguide individually at the linear array end, and the transmitted light was collected at the MCF coupling end using a highly MM, 600 µm core diameter fiber directly butt-coupled to the fan-out. This large core diameter fiber was used in order to capture all the light guided by the coupled waveguide array. The light collected by the MM fiber was measured at the other end of the MM fiber using a germanium detector. Index matching gel was used at all fiber-waveguide interfaces in order to reduce Fresnel reflections. For our purposes, the insertion loss of each waveguide was defined as the difference in signal measured when the SMF-28 fiber and MM fiber were directly coupled to each other, compared to when they were each coupled to the fan-out.

Fig. 3 presents the results of the insertion losses measurements. It can be seen that the insertion losses of the waveguides vary between 1.6 dB and 3.26 dB. Summing across the measurements for all waveguides, the fan-out exhibits a total throughput loss of 2.0 dB. Using the Gaussian field approximation [14], the minimum coupling loss due to mode-mismatch between the fan-out waveguides and the SMF-28 fiber (which exhibits a MFD of 10.4 ± 0.8 µm) is ≈ 0.35 ± 0.15 dB. These results indicate that if negligible coupling losses could be achieved between the MCF and the fan-out, this fan-out would exhibit a total throughput loss of ≈ 1.7 dB.

To connect the MCF to the fan-out, the MCF was glued into a glass V-groove and polished back. To align the MCF with the fan-out, we first illuminated all cores of the MCF using white light coupled into the MCF at the opposite end from the V-groove. We then butt-coupled the MCF+V-groove to the fan-out, and viewed the light transmitted by the fan-out as the rotation and position of the MCF was adjusted. Once an apparently optimal position of the MCF had been achieved, we secured the MCF in place using UV curing epoxy.

The insertion loss of the fan-out+MCF combination was tested by coupling light into each fan-out waveguide at the linear array end. We defined the insertion loss in this case as the difference in signal measured when the input coupling SMF-28 fiber was coupled directly to the detector, compared to when the light was coupled into the fan-out and the light emerging from the unconnected end of the MCF was measured. The results of these measurements are also shown in Fig. 3, where it can be seen that the insertion loss of the fan-out+MCF varies considerably between 5.0 and 13.9 dB. Summing across the measurements for all waveguide inputs, we calculate that the fan-out+MCF combination exhibits an overall throughput loss of 7.0 dB.

Clearly, the coupling losses between the fan-out and the MCF are responsible for the lowered throughput of the fan-out+MCF combination. The MCF cores were known to support rotationally symmetric single guided modes with a 1/e$^2$ diameter of ≈ 6.6 µm. Again, using the Gaussian field approximation [14], we estimate that the minimum coupling loss due to mode-mismatch between the fan-out waveguides and the MCF cores is ≈ 0.6 dB. The mode-mismatch is, therefore, insufficient on its own to explain the lowered throughput of the fan-out+MCF combination, and we conclude that alignment accuracy between the fan-out waveguides and the MCF cores is primarily responsible. From Fig. 3, for example, it can be seen that 23 of the cores exhibit an insertion loss > 10 dB. Of these 23 cores, 22 are in the bottom (lowest z position) four rows of cores in the MCF coupling end of the fan-out.

This is clear evidence that spatial alignment error is primarily responsible for the low throughput, which can be readily addressed in the future through the implementation of improved alignment techniques.

Finally, for the final OH-line suppression application we require total throughput losses < 1.0 dB. Currently, the individual waveguides are close to achieving this level of performance, but the alignment between the MCF and the fan-out must be improved significantly. To conclude, we have demonstrated that ultrafast laser inscribed fan-out devices are a promising way for reformatting the output of MCF-PLs into a one-dimensional array of single modes. The full development of such devices is therefore crucial to future applications of mass-producible MCF-PLs in astronomy.

### Acknowledgments
This work was funded by the UK – STFC through RRT's Advanced Fellowship (ST/H005595/1) and by the UK – EPSRC (EP/G030227/1). TAB thanks the Leverhulme Trust for a Research Fellowship, Brian Mangan for help in fiber fabrication and the Astrophysikalisches Institut Potsdam for providing materials. JBH is supported by an Australian Research Council Federation Fellowship. RRT thanks A. K. Kar for access to the ULI facility.

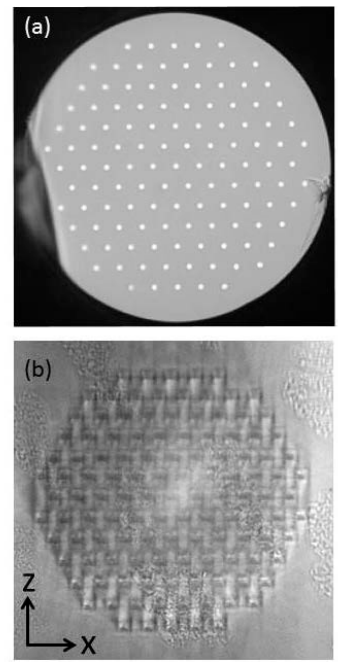

Fig. 1 Optical micrographs of (a) The cleaved end-facet of the 120 core multicore fiber. (b) The multicore fiber coupling end of the three-dimensional waveguide fan-out.

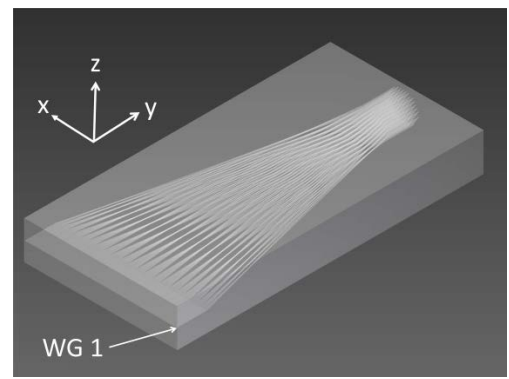

Fig. 2 Sketch of the three-dimensional waveguide fan-out device. The position of waveguide 1 (WG1) is indicated in the sketch. The waveguide number then increases sequentially along the linear array.

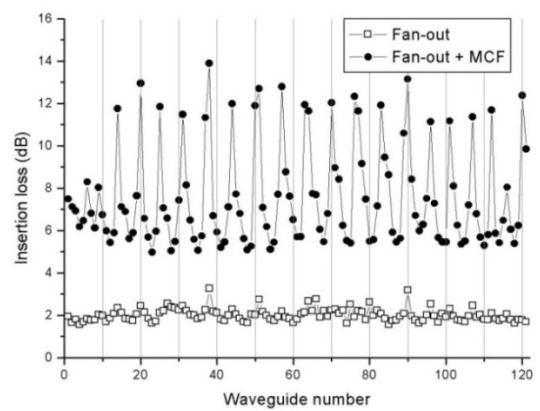

Fig. 3 Insertion losses measured for the individual fan-out waveguides and the fanout+MCF combination.